\documentclass[prl,amsfonts,amssymb,floats,twocolumn,superscriptaddress,aps]{revtex4}

\usepackage[final,dvips]{epsfig}

\begin{document}

\title[]
{
Quantum phase transitions in the sub-ohmic spin-boson model: \\
Failure of the quantum--classical mapping
}
\author{Matthias Vojta}
\affiliation{\mbox{Institut f\"ur Theorie der Kondensierten Materie,
Universit\"at Karlsruhe,
76128 Karlsruhe, Germany}}
\author{Ning-Hua Tong}
\affiliation{\mbox{Theoretische Physik III, Elektronische Korrelationen und
Magnetismus, Universit\"at Augsburg, Germany}}
\author{Ralf Bulla}
\affiliation{\mbox{Theoretische Physik III, Elektronische Korrelationen und
Magnetismus, Universit\"at Augsburg, Germany}}
\date{Jan 20, 2005}

\begin{abstract}
The effective theories for many quantum phase transitions can be mapped
onto those of classical transitions.
Here we show that the naive mapping fails for the sub-ohmic spin-boson model
which describes a two-level system coupled to a bosonic bath
with power-law spectral density, $J(\omega)\propto\omega^s$.
Using an $\epsilon$ expansion we prove that this model has a quantum transition
controlled by an {\em interacting} fixed point at small $s$, and support this
by numerical calculations.
In contrast, the corresponding classical long-range Ising model is known to
display mean-field transition behavior for $0<s<1/2$,
controlled by a {\em non-interacting} fixed point.
The failure of the quantum--classical mapping is argued to arise from the long-ranged
interaction in imaginary time in the quantum model.
\end{abstract}

\pacs{PACS: 05.30.Cc (Renormalization Group methods),
05.30.Jp (Boson systems)}

\maketitle


Low-energy theories for certain classes of quantum phase transitions
in clean systems with $d$ spatial dimensions are known to be equivalent to the
ones of classical phase transitions in $(d+z)$ dimensions,
where $z$ ist the dynamical exponent of the quantum transition \cite{book}.
This mapping is usually established in a path integral formulation
of the effective action for the order parameter,
where imaginary time in the quantum problem takes
the role of $z$ additional space dimensions in the classical
counterpart.
The tuning parameter for the phase transition, being the ratio of certain
coupling constants in the quantum problem (where $T$ is fixed to zero),
becomes temperature for the classical transition.
For the quantum Ising model, where the transverse field
can drive the system into a disordered phase at $T=0$, the quantum--classical
equivalence in the scaling limit can be explicitly shown using transfer
matrix techniques \cite{book}.
While this formal proof is only applicable for {\em short-range} interactions in time
direction, it is believed that it also holds for long-range interactions,
which can arise upon integrating out gapless degrees of freedom coupled to
the order parameter.
(Counter-examples are phase transitions in itinerant magnets,
where the elimination of low-energy fermions produces non-analyticities in
the resulting order parameter field theory \cite{bkv}.)
A paradigmatic example is the spin-boson model \cite{Leggett, Weiss},
where an Ising spin
(i.e. a generic two-level system) is coupled to a bath of harmonic
oscillators: eliminating the bath variables leads to a retarded
self-interaction for the local spin degree of freedom, which decays as
$1/\tau^2$ in the well-studied case of ohmic damping.
Interestingly, the same model is obtained as the low-energy limit
of the anisotropic Kondo model which describes a spin-1/2 magnetic
impurity coupled to a gas of conduction electrons \cite{yuval,emery}.

The purpose of this paper is to point out that the naive quantum--classical
mapping can fail for long-ranged interactions in imaginary time even for
the simplest case of $(0+1)$ dimensions and Ising symmetry.
We shall explicitly prove this failure for the sub-ohmic spin-boson model,
by showing that the phase transitions in the quantum problem and in the corresponding
classical long-range Ising model fall in different universality classes.

The spin-boson model is described by the Hamiltonian
\begin{equation}
{\cal H}_{\rm SB}=-\frac{\Delta}{2}\sigma_{x}+\frac{\epsilon}{2}\sigma_{z}+
\sum_{i} \omega_{i}
     a_{i}^{\dagger} a_{i}
+\frac{\sigma_{z}}{2} \sum_{i}
    \lambda_{i}( a_{i} + a_{i}^{\dagger} )
\label{eq:sbm}
\end{equation}
in standard notation.
The coupling between spin $\sigma$ and the bosonic bath with oscillators $\{a_i\}$
is completely specified by the bath spectral function
\begin{equation}
    J\left( \omega \right)=\pi \sum_{i}
\lambda_{i}^{2} \delta\left( \omega -\omega_{i} \right) \,,
\end{equation}
conveniently parametrized as
\begin{equation}
  J(\omega) = 2\pi\, \alpha\, \omega_c^{1-s} \, \omega^s\,,~ 0<\omega<\omega_c\,,\ \ \ s>-1
\label{power}
\end{equation}
where the dimensionless parameter $\alpha$ characterizes the
dissipation strength, and $\omega_c$ is a cutoff energy.
The value $s=1$ represents the case of ohmic dissipation,
where a Kosterlitz-Thouless transition separates a delocalized phase
at small $\alpha$ from a localized phase at large $\alpha$.
These two phases asymptotically correspond to eigenstates of
$\sigma_x$ and $\sigma_z$, respectively.

In the following, we are interested in sub-ohmic damping, $0<s<1$
\cite{spohn,KM}.
The standard approach is to integrate out the bath, leading to an
effective interaction
\begin{eqnarray}
{\cal S}_{\rm int} = \int d\tau d\tau' \sigma_z(\tau) g(\tau-\tau') \sigma_z(\tau')
\end{eqnarray}
with $g(\tau) \propto 1/\tau^{1+s}$ at long times.
Numerical renormalization group (NRG) calculations in Refs.~\onlinecite{BTV,BLTV},
performed directly for the sub-ohmic spin-boson model,
have established that a
second-order quantum transition occurs for all $0<s<1$.
Here we use an analytical renormalization group (RG) expansion,
controlled by the small parameter $s$,
to establish that the spin-boson transition at small $s$ is governed by
an interacting fixed point with strong hyperscaling properties.
This analytical result is supported by NRG calculations.
In contrast, the transition in the classical Ising model is known to display
mean-field behavior for $0<s<1/2$ \cite{fisher,luijten}.


{\em Scaling and critical exponents.}
A scaling ansatz for the impurity part of the free energy takes the form
\begin{equation}
F_{\rm imp} = T f(|\alpha-\alpha_c| T^{-1/\nu}, \epsilon T^{-y_\epsilon} )
\label{fscal}
\end{equation}
where $|\alpha-\alpha_c|$ measures the distance to criticality.
The bias $\epsilon$ takes the role of a local field (with scaling exponent $y_\epsilon$);
and $\nu$ is the correlation length exponent which describes the vanishing of the energy scale
$T^\ast$, above which critical behavior is observed \cite{book}:
$T^\ast \propto |\alpha-\alpha_c|^{\nu}$.
The ansatz (\ref{fscal}) assumes the fixed point to be interacting;
for a Gaussian fixed point the scaling function will also depend upon
dangerously irrelevant variables.

With the local magnetization
$M_{\rm loc} = \langle\sigma_z\rangle = -\partial F_{\rm imp}/\partial\epsilon$
and the susceptibility
$\chi_{\rm loc} = -\partial^2 F_{\rm imp}/(\partial\epsilon)^2$
we can define critical exponents (see also Ref.~\onlinecite{insi}):
\begin{eqnarray}
M_{\text{loc}}(\alpha > \alpha_c,T=0,\epsilon=0)
&\propto& (\alpha-\alpha_c)^{\beta}, \nonumber\\
\chi_{\text{loc}}(\alpha < \alpha_c,T=0) &\propto& (\alpha_c-\alpha)^{-\gamma},
\nonumber\\[-1.75ex]
\label{exponents} \\[-1.75ex]
M_{\text{loc}}(\alpha=\alpha_c,T=0) &\propto& | \epsilon |^{1/\delta}, \nonumber\\
\chi_{\text{loc}}(\alpha=\alpha_c,T) &\propto&
T^{-x}, \nonumber \\
\chi_{\text{loc}}''(\alpha=\alpha_c,T=0,\omega) &\propto&
|\omega|^{-y} {\rm sgn}(\omega). \nonumber
\end{eqnarray}
The last equation describes the dynamical scaling of $\chi_{\rm loc}$. 

In the absence of a dangerously irrelevant variable there are only two independent
exponents, e.g., $\nu$ and $y_\epsilon$.
The scaling form (\ref{fscal}) yields hyperscaling relations:
\begin{eqnarray}
\beta = \gamma \frac{1-x}{2x},~~
2\beta + \gamma = \nu, ~~
\gamma = \nu x,~~
\delta = \frac{1+x}{1-x} \,.
\end{eqnarray}
Hyperscaling also implies $x=y$, which is equivalent to
so-called $\omega/T$ scaling in the dynamical behavior.


{\em Long-range Ising model.}
The classical counterpart of the spin-boson model (\ref{eq:sbm}) is the
one-dimensional Ising model \cite{Leggett, Weiss}
\begin{equation}
{\cal H}_{\rm cl} = - \sum_{\langle ij \rangle} J_{ij} S_i^z S_j^z + {\cal H}_{\rm SR}
\label{hcl}
\end{equation}
with interaction $J_{ij} = J/|i-j|^{1+s}$.
${\cal H}_{\rm SR}$ contains an additional generic short-range interaction
which arises from the transverse field, but is believed to be irrelevant
for the critical behavior \cite{fisher,luijten}.
As proven by Dyson \cite{dyson} this model displays a phase transition for $0<s\leq 1$.
Both analytical arguments, based on the equivalence to a O(1) $\phi^4$ theory \cite{fisher},
and extensive numerical simulations \cite{luijten}
show that the upper-critical dimension for the $d$-dimensional long-range
Ising model is $d_c^+ = 2s$, i.e., in $d=1$ the transition obeys
non-trivial critical behavior for $1/2<s<1$.
In contrast, mean-field behavior obtains for $0<s<1/2$, with exponents \cite{fisher,luijten}
$\beta = 1/2$, $\gamma = 1$, $\delta = 3$, $\nu = 1/s$, $y=s$
violating hyperscaling.

As the exponent $s$ exclusively determines the power laws of spectra and
correlations in a long-range model once spatial dimensionality ($d=1$) is fixed,
$s$ takes the role of a ``dimension'', i.e., we will refer to $s=1/2$ as
upper-critical ``dimension'' of the classical Ising chain.

Near $s=1$ the phase transition can be analyzed
using a kink gas representation of the partition function,
where the kinks represent Ising domain walls \cite{koster}.
This expansion, controlled by the smallness of the kink fugacity, is done around
the ordered phase of the Ising model, corresponding to the localized fixed point
of the spin-boson quantum problem.
For small $(1-s)$ the results obtained via the perturbative kink-gas RG
are consistent with the NRG data for the spin-boson transition \cite{BTV},
indicating that the quantum--classical
mapping works in the asymptotic vicinity of the localized fixed point.


{\em Spin-boson model: perturbative RG.}
We now describe a novel RG expansion which is performed around
the {\em delocalized} fixed point of the spin-boson model.
NRG indicates that the critical fixed point merges 
with the delocalized one as $s\to 0^+$, thus we expect that the expansion
will allow access to the quantum phase transition for
small $s$.
As shown below, the expansion is done about the {\em lower-critical}
``dimension'' $s=0$; it yields an interacting fixed point, and mean-field critical
behavior for small $s$ does {\em not} obtain.
For convenience we assume equal couplings, $\lambda_i \equiv \lambda$,
then the energy dependence of $J(\omega)$ is contained in the density of states of
the oscillator modes $\omega_i$, and we have $\alpha \propto \lambda^2$.

How to set up a suitable RG expansion?
Power counting about the free-spin fixed point, $\lambda=\Delta=0$, gives the scaling dimensions
${\rm dim}[\lambda] = (1-s)/2$, ${\rm dim}[\Delta] = 1$.
Thus, both parameters are strongly relevant for small $s$.
A better starting point is the delocalized fixed point, corresponding to
{\em finite} $\Delta$.
Eigenstates of the impurity are $|\!\rightarrow_x\rangle$ and $|\!\leftarrow_x\rangle$,
with an energy splitting of $\Delta$.
The low-energy Hilbert space contains the state $|\!\rightarrow_x\rangle$
only, and interaction processes with the bath arise in second-order
perturbation theory, proportional to $\kappa_0 = \lambda^2/\Delta$.
Power counting w.r.t. the $\lambda=0$ limit now gives
${\rm dim}[\kappa_0] = -s$, i.e., $\kappa_0$ is marginal
at $s=0$, indicating that an $\epsilon$-type expansion for small $s$
is possible.

\begin{figure}[!t]
\epsfxsize=3.1in
\centerline{\epsffile{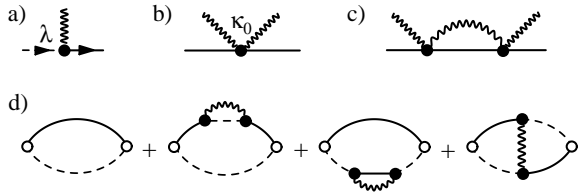}}
\caption{
Feynman diagrams occurring in the perturbation theory for the
spin-boson model.
Full/dashed lines denote the propagators of the $|\!\rightarrow_x\rangle$ and $|\!\leftarrow_x\rangle$
impurity states -- the two states are separated by a gap of size $\Delta$.
The wiggly line is the local bath boson $G_{\rm loc}$.
a) Interaction vertex $\lambda$.
b) Interaction vertex $\kappa_0$ in the low-energy sector.
c) One-loop renormalization of $\kappa$.
d) Diagrams for the local susceptibility $\chi_{\rm loc}$.
}
\label{fig:dgr}
\end{figure}

To lowest order, the RG can be performed within the low-energy sector,
i.e., for the $\kappa_0$ vertex (Fig.~\ref{fig:dgr}b) and the propagator
for the $|\!\rightarrow_x\rangle$ state.
Consequently, our approach is valid as long as $\lambda \ll \Delta, \omega_c$.
We introduce a renormalized coupling $\kappa$ according to
$\kappa_0 = \Lambda^{-s} \kappa$ where $\Lambda$ is the running cutoff,
$\Lambda=\omega_c$ initially.
The one-loop beta function can be derived using the familiar momentum shell method,
i.e., by successively eliminating high-energy bath bosons.
To one-loop order only the coupling-constant renormalization in Fig.~\ref{fig:dgr}c
enters, and we obtain
\begin{eqnarray}
\beta(\kappa) = -s \kappa + \kappa^2 \,.
\end{eqnarray}
Besides the stable delocalized fixed point $\kappa=0$ this
flow equation displays an infrared unstable fixed point at
\begin{equation}
\kappa^\ast = s + {\cal O}(s^2)
\label{fp}
\end{equation}
which controls the transition between the delocalized and
localized phases.
No (dangerously) irrelevant variables are present in this theory,
so we conclude that the critical fixed point (\ref{fp}) is interacting
\cite{potnote}.

We proceed with the calculation of critical exponents.
Expanding the RG beta function around the fixed point (\ref{fp})
gives the correlation length exponent
\begin{equation}
1 / \nu = s + {\cal O}(s^2) \,,
\label{nu}
\end{equation}
i.e., $\nu$ diverges as $s\to 0^+$, as characteristic for a lower-critical dimension.
Parenthetically, we note that the RG structure for $s\to 1^-$ is also similar to that
near a lower-critical dimension:
The line of second-order transitions for $0<s<1$ terminates in a Kosterlitz-Thouless
transition at $s=1$ and is thus bounded by {\em two lower-critical ``dimensions''} --
a similar situation was recently found in the pseudogap Kondo problem, which, however,
is in a different universality class \cite{LFMV}.
As usual for RG expansions around a lower-critical dimension the present RG can
only capture one of the two phases (the delocalized one),
whereas run-away flow occurs on the localized side.

The exponents associated with the local field $\epsilon$ can be obtained
in straightforward renormalized perturbation theory.
To calculate observables, the diagrams are written down using
the original model with couplings $\lambda$ and $\Delta$ and
both $|\!\rightarrow_x\rangle$, $|\!\leftarrow_x\rangle$ states.
The perturbation theory turns out to be organized in powers of $\lambda^2/\Delta$,
as expected.
Some of the relevant diagrams are displayed in Fig.~\ref{fig:dgr}d,
details will appear elsewhere.
Restricting ourselves to the lowest-order results for the disordered and
quantum critical regimes we find
\begin{eqnarray}
\gamma &=& 1 + {\cal O}(s) ,~~ x = y = s + {\cal O}(s^2), \nonumber\\
1/\delta &=& 1 - 2 s + {\cal O}(s^2) \,.
\label{delta}
\end{eqnarray}
Interestingly, we are able to derive an {\em exact} result for the exponents
$x$, $y$, employing an argument along the lines of Refs.~\onlinecite{vbs,MVMK},
based on the diagrammatic structure of $\chi_{\rm loc}$.
We obtain
\begin{eqnarray}
x = y = s \label{x} \,.
\end{eqnarray}
(Notably, $y=s$ was found to be the exact decay exponent of the critical
spin correlations in the long-range Ising model for all $s$ \cite{fisher,suzuki}.)
Hyperscaling yields $\delta = (1+s)/(1-s)$,
consistent with the lowest-order result (\ref{delta}).

\begin{figure}[!t]
\epsfxsize=3.3in
\centerline{\epsffile{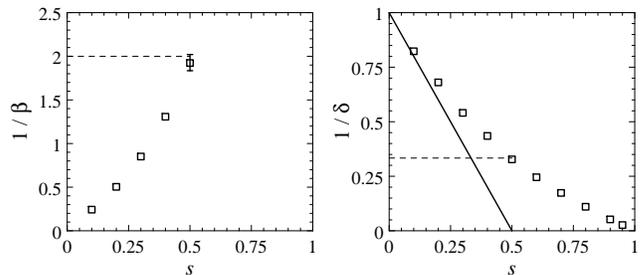}}
\caption{
Left:
NRG data for the order parameter exponent $1/\beta$.
Right:
Magnetization exponent $1/\delta$ from NRG,
together with the analytical RG result (\ref{delta}) [solid line].
The dashed lines are the mean-field results $\beta = 1/2$, $\delta=3$.
}
\label{fig:betadelta}
\end{figure}

As an aside, we note that at $s=0$ the bath coupling is marginally relevant.
Therefore the impurity is always localized as $T\to 0$, with a localization
temperature given by $T^\ast = \omega_c \exp(-\Delta\omega_c / \lambda^2)$.


\begin{figure}[b]
\epsfxsize=3.3in
\centerline{\epsffile{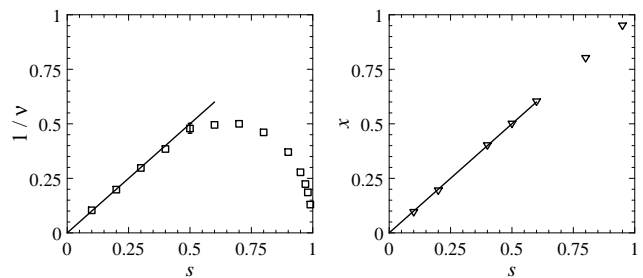}}
\caption{
Left:
Numerical data for the correlation length exponent $1/\nu$ obtained from NRG,
together with the RG result (\ref{nu}).
Right:
Susceptibility exponent $x$ from NRG,
together with the RG result (\ref{x}).
Here, the mean-field results coincide with the lowest-order perturbative ones.
}
\label{fig:nux}
\end{figure}

{\em Spin-boson model: Numerical results.}
We have performed extensive NRG calculations \cite{BLTV}
to evaluate the critical exponents of the spin-boson transition,
see Refs.~\onlinecite{BTV,BLTV} for numerical
details \cite{footnrg}.
Results are shown in Figs.~\ref{fig:betadelta}
and \ref{fig:nux}: these exponents obey hyperscaling
including $x=y$ ($\omega/T$ scaling).
They are in excellent agreement with the small-$s$ RG expansion,
but at variance with the exponents of the long-range Ising
model: the mean-field predictions are $\beta = 1/2$, $\delta = 3$
which are clearly violated by our results in Fig.~\ref{fig:betadelta}.

Within error bars the mean-field exponents are realized {\em at}
$s=1/2$.
Further, the one-loop results for $\nu$ (\ref{nu}) and $\gamma$ (\ref{delta})
appear to be exact for all $0<s<1/2$,
and logarithmic corrections to the power laws are observed at $s=1/2$.
This suggests that the spin-boson transition {\em does} change its character at
$s=1/2$,
but the critical fixed points for both $0<s<1/2$ and $1/2<s<1$ are interacting
and obey hyperscaling
(the latter one being equivalent to that of the classical Ising model).


{\it Discussion.}
We have proven that the naive quantum--classical mapping fails for the
sub-ohmic spin-boson model:
Using a novel RG expansion around the delocalized fixed point, we have
shown that the quantum transition at $0<s<1/2$ is controlled by an
interacting fixed point, whereas the corresponding classical long-range
Ising model shows mean-field behavior.
Thus, the spin-boson problem for $s<1/2$ is equivalent neither to
the classical Ising model nor to the corresponding (quantum or classical)
O(1) $\phi^4$ theory \cite{luijten}.
In physical terms the inequivalence can be traced back to the different
disordered (delocalized) fixed points in the two situations
(expansions around these fixed points are suitable to access the critical
behavior for small $s$):
In the quantum model the transverse field fully polarizes the spin in $x$ direction
(which can be viewed as a ``condensate'' of spin flips),
whereas the high-temperature limit of the classical Ising model is simply
incoherently disordered.

The inequivalence may come unexpected -- so where does the quantum--classical mapping fail?
Formal proofs of the mapping using transfer matrices
rely on the short-ranged character of the interaction \cite{book}.
For general interactions a Trotter decomposition of the quantum partition
function is employed where the imaginary axis of length
$\beta = 1/T$ is divided into $N$ slices of size $\Delta\tau=\beta/N$,
leading to an Ising chain (\ref{hcl}) with $N$ sites.
This procedure is exact when the limits $\Delta\tau\to 0$ and $\beta\to\infty$
are taken in this order.
However, the limit $\Delta\tau\to 0$ leads to a {\em diverging} near-neighbor coupling
in the term ${\cal H}_{\rm SR}$ of the classical Ising model (\ref{hcl}) \cite{emery}.
This may in fact change the critical behavior of ${\cal H}_{\rm cl}$ (\ref{hcl}),
as a classical model with finite couplings arises upon taking $\beta\to\infty$ first.
In other words, the quantum and classical problems are only equivalent
if and only if the low-energy limit of ${\cal H}_{\rm cl}$ is independent
of the order of the two limits $\Delta\tau\to 0$ and $\beta\to\infty$ \cite{emery}.
As we have proven the inequivalence of ${\cal H}_{\rm SB}$ and ${\cal H}_{\rm cl}$
(with finite couplings) we conclude that the two limits cannot be interchanged for $s<1/2$.

A recent paper \cite{stefan} investigated a SU($N$)-symmetric Bose-Fermi Kondo model
in a certain large-$N$ limit and found a critical fixed point with $\omega/T$
scaling for all $s$.
The authors argued that the apparent failure of the quantum--classical mapping is
due to the quantum nature of the impurity spin.
As discussed above, this alone is not sufficient:
for short-range interactions in imaginary time the mapping can be proven to be
asymptotically exact \cite{book};
{\em long-range} interactions are essential.

Our results suggest that some conclusions drawn in the past
for effectively long-range Ising systems on the basis of the
quantum--classical mapping have to be re-examined.
Further, we envision that our novel $\epsilon$ expansion will have applications
for various quantum impurity problems, e.g., two-level systems coupled to
multiple baths, Bose-Fermi Kondo models \cite{MVMK,stefan},
and also for quantum dissipative lattice models \cite{lattice}.

We thank D. Gr\"uneberg, H. Rieger, A. Rosch, S. Sachdev,
Q. Si, M. Troyer, T. Vojta, and W. Zwerger for discussions.
This research was supported by the DFG through SFB 484 (RB) and
the CFN Karlsruhe (MV), and by
the Alexander von Humboldt foundation (NT).


\newpage

\begin{widetext}
\begin{center}
~\\
{\large\bf Erratum: Quantum phase transitions in the sub-ohmic spin-boson model:
Failure of the quantum--classical mapping,
Phys. Rev. Lett. {\bf 94}, 070604  (2005)}\\
~\\
Matthias Vojta, Ning-Hua Tong, and Ralf Bulla\\
~\\
We point out that the interpretation of data obtained by the Numerical Renormalization Group\\
(NRG) method in Phys. Rev. Lett. {\bf 94}, 070604 (2005) was incorrect: Two different effects precluded\\
the reliable determination of critical exponents. An analytical argument was as well incomplete.\\
We discuss consequences for the quantum phase transition scenario.\\
~\\
\end{center}
\end{widetext}


Ref. \onlinecite{prl} used results from NRG to argue that
the critical properties of the spin-boson model with bath exponent $s<1/2$
are not those of mean-field theory.
The difference is in the order-parameter exponents
$\beta$, $\delta$ and $x$.

By now, we have convinced ourselves that the exponents in question
could not be reliably extracted from the NRG calculations,
if the ordered phase is controlled by an operator which is
dangerously irrelevant at criticality.
The two relevant issues will be discussed in turn.

(A) {\it Hilbert-space truncation}:
The bosonic NRG is presently unable to describe
the physics of the localized fixed point for $s<1$, due to the truncation of
the bosonic Hilbert space \cite{BLTV2}.
From the absence of convergence problems at the critical fixed point
we nevertheless concluded \cite{prl} that all critical exponents can be extracted
from NRG.
While correct for an interacting fixed point, this assertion is erroneous
for exponents $\beta$ and $\delta$ of a Gaussian fixed point,
because, in the presence of a dangerously irrelevant variable,
those exponents are {\em not} properties of the critical fixed point and its
vicinity, but instead properties of the flow towards the fixed point controlling
the ordered phase.
(This is what happens above the upper-critical dimension of a standard $\phi^4$ theory.)

(B) {\it Mass flow:}
The NRG algorithm integrates out the bath iteratively. In the order-parameter language,
this leads to a mass renormalization along the flow, arising from the bath (not from
non-linear interactions), i.e.,
at any NRG step only a partial mass renormalization has been taken into account.
As a result, the critical flow trajectory in NRG is that of a system with an additional
mass set by the current NRG scale (i.e. by the temperature $T$),
and, effectively, the system is not located at the critical coupling for any finite $T$.
For an interacting fixed point, the missing mass correction from the bath and
the interaction-generated mass have the same scaling, hence no qualitative error is
introduced.
In contrast, above the upper-critical dimension, the interaction-generated
mass vanishes faster as $T\!\to\!0$,
and the bath mass correction dominates all observables calculated along
the critical flow trajectory, like the exponent $x$.
Note that this conceptual ``mass-flow'' problem of NRG arises for all baths without
low-energy particle-hole symmetry.

Consequently, the NRG-extracted exponents $\beta$, $\delta$, and $x$
quoted in Ref.~\onlinecite{prl} were unreliable.
The same applies to the analytical argument given
in Ref.~\onlinecite{prl}: The RG equations (9-11) were correct, but the
following analysis relied on the absence of dangerously irrelevant
variables as well.

Analyzing (A) in detail, we have found that careful NRG calculations performed
with a large number $N_b$ of bosonic states in principle allow to extract
the correct exponents $\beta$ and $\delta$.
While the truncation always dominates asymptotically close
to criticality, the physical power laws can be seen on intermediate scales.
As an example, we show in Fig.~1 the non-linear field response at criticality for
$s=0.4$. For $N_b=50$ a crossover from $\delta\approx2.4$ to the mean-field value $\delta=3$
upon increasing the field is visible, before the quantum critical regime is left.

\begin{figure}[b]
\includegraphics[width=3.5in]{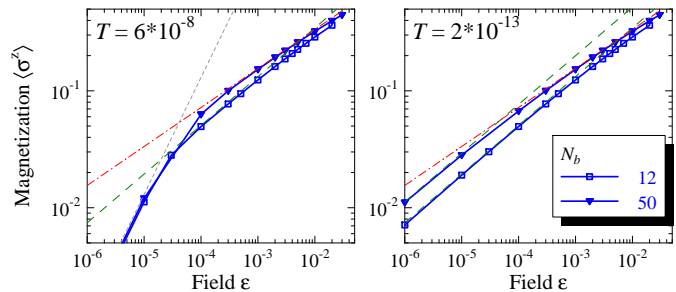}
\caption{
Magnetization curve of the critical spin-boson model for $s=0.4$.
NRG parameters are $\Lambda=4$, $(N_s,N_b)=(40,12)$ and $(60,50)$ \cite{BLTV2}.
The lines are power laws with exponents 1 (dotted), 1/2.4 (dashed), and 1/3 (dash-dot).
\vspace{-15pt}
}
\end{figure}

Taking into account recent numerical work based on Quantum Monte Carlo
and exact diagonalization techniques \cite{rieger,fehske},
which confirmed mean-field behavior in the sub-ohmic spin-boson model for $s<1/2$,
we conclude that the quantum-to-classical mapping is valid here.



\begin{references}{}
\bibitem{potnote}
In the low-energy sector, $|\!\rightarrow_x\rangle$, the effective theory reduces to a potential
scattering problem for the bath bosons, which can be solved exactly,
and indeed has a phase transition at $\kappa_0 G_{\rm loc}(\omega=0) = 1$.
Corrections can arise from higher-order processes
involving $|\!\leftarrow_x\rangle$.
\bibitem{footnrg}
Within NRG the boson occupation number stays small for
all $s$ in the vicinity of the critical fixed points
(in contrast to the localized fixed points, where it diverges in
the low-energy limit).
Thus, the bosonic chain-NRG \protect\cite{BLTV} can be used which has been proven to give
reliable results in the quantum critical regime, e.g., for exponents.
\end{references}
\end{document}